\documentclass[12pt]{article}
\usepackage{graphicx}
\usepackage{epsfig}
\usepackage{epstopdf}
\DeclareGraphicsExtensions{.pdf,.eps,.png,.jpg,.mps}

\def\lsim{\mathrel{\rlap{\lower3pt\hbox{\hskip0pt$\sim$}}
     \raise1pt\hbox{$<$}}}         
\def\gsim{\mathrel{\rlap{\lower4pt\hbox{\hskip1pt$\sim$}}
     \raise1pt\hbox{$>$}}}         

\usepackage{amsmath}
\usepackage{amsfonts}

\begin{document}
\begin{titlepage}

{\centerline{\Large \bf Non-perturbative Massive de Sitter Solutions}}
\medskip

\centerline{Zurab Kakushadze$^\S$$^\dag$$^\ddag$\footnote{\, Email: \tt zura@quantigic.com}}
\bigskip

\centerline{\em $^\S$ Quantigic$^\circledR$ Solutions LLC}
\centerline{\em 1127 High Ridge Road \#135, Stamford, CT 06905\,\,\footnote{\, DISCLAIMER: This address is used by the corresponding author for no
purpose other than to indicate his professional affiliation as is customary in
publications. In particular, the contents of this paper
are not intended as an investment, legal, tax or any other such advice,
and in no way represent views of Quantigic® Solutions LLC,
the website \underline{www.quantigic.com} or any of their other affiliates.
}}
\centerline{\em $^\dag$ Department of Physics, University of Connecticut}
\centerline{\em 1 University Place, Stamford, CT 06901}
\centerline{\em $^\ddag$ Free University of Tbilisi, Business School \& School of Physics}
\centerline{\em 240, David Agmashenebeli Alley, Tbilisi, 0159, Georgia}
\medskip
\centerline{(May 15, 2014; revised: October 6, 2014)}

\bigskip
\medskip

\begin{abstract}
{}We discuss non-perturbative dynamics of massive gravity in de Sitter space via gravitational Higgs mechanism.
We argue that enhanced local symmetry and null (ghost) state at (below) the perturbative Higuchi bound are mere artifacts of not only linearization
but also assuming the Fierz-Pauli mass term. We point out that, besides de Sitter, there are vacuum solutions where the space asymptotically is de Sitter both in the past and in the future, the space first contracts, this contraction slows down, and then reverses into expansion, so there is an epoch where the space appears to be (nearly) flat, even though the vacuum energy density is non-vanishing. We confirm this by constructing a closed-form exact solution to full non-perturbative equations of motion for a ``special" massive de Sitter case. We give a formula for the ``critical" mass above which such solutions apparently do not exist. For the Fierz-Pauli mass term this ``critical" mass coincides with the perturbative Higuchi bound, and the former serves as the non-perturbative reinterpretation of the latter. We argue that, notwithstanding the perturbative ghost, non-perturbatively there is no ``instability". Instead, there are additional vacuum solutions that may have interesting cosmological implications, which we briefly speculate on.

\end{abstract}
\end{titlepage}

\newpage

\section{Introduction and Summary}

{}The currently observed accelerated expansion of the Universe \cite{Nova,Nova1} is one of the motivations for considering large-scale modifications of gravity. One such modification is giving small mass to the graviton with the aim of explaining the accelerated expansion without a tiny cosmological constant.\footnote{\, Notwithstanding that zero cosmological constant might be just as ``unnatural".} While -- barring unbroken supersymmetry -- the cosmological constant is not protected, quantum corrections to the graviton mass appear to be suppressed by the graviton mass itself \cite{dR}. However, unlike in a scalar or vector theory, in gravity there is more to the graviton mass.

{}A general Lorentz invariant mass term for the graviton $h_{MN}$ in the linearized approximation is of the form
\begin{equation}\label{mass.term}
 -{M^2\over 4} \left[h_{MN}h^{MN} - \beta (h^M_M)^2\right]~,
\end{equation}
where $\beta$ is a dimensionless parameter. Perturbatively,\footnote{\, Here we refer to perturbative expansion in powers of $h_{MN}$ in the classical theory, not loop corrections.}  for $\beta\neq 1$ the trace component $h^M_M$ is a propagating ghost, while
it decouples in the Minkowski background for the Fierz-Pauli mass term with $\beta = 1$ \cite{FP}. Again, perturbatively, generically a ghost reappears beyond the linearized level even for the Fierz-Pauli mass term \cite{BD} and, once again, perturbatively, to avoid the reappearance of a ghost beyond the linearized level one attempts tuning the mass term to a special form \cite{dR}. On the other hand, while quantum mechanically small graviton mass $M$ {\em per se} may well be ``technically natural" \cite{dR}, the special form of the mass term necessary for avoiding a ghost beyond the linearized level is not protected by any symmetry. In fact, already at the linearized level $\beta$ in (\ref{mass.term}) is not protected by any symmetry against quantum corrections. Does this then mean that massive gravity is ``unstable"?

{}Gravitational Higgs mechanism \cite{KL, thooft} provides a non-perturbative and fully covariant definition of massive gravity. Non-perturbatively, even for $\beta\neq 1$, the Hamiltonian is bounded from below and the perturbative ghost is merely an artifact of linearization \cite{dS, Unitarity}.\footnote{\, The full non-perturbative Hamiltonian for the model of \cite{thooft}, which has $\beta=1/2$, in the gravitational Higgs mechanism framework was constructed in \cite{JK} and is expressly positive-definite. Non-perturbative unitarity for general $\beta$ in the Minkowski space was argued in \cite{Unitarity}, while for the de Sitter space it was argued in \cite{dS}. For a list of other works related to gravitational Higgs mechanism and massive gravity, see \cite{ZK.massive} and \cite{dR}.} Therefore, stability should be addressed in the context of full non-perturbative theory. Furthermore, since there is no symmetry that would protect $\beta$, there appears to be no reason to restrict to $\beta=1$. In particular, if there is any ``instability", it should be visible non-perturbatively. For instance, in \cite{ZK.massive}, in furtherance of the results of \cite{Unitarity}, explicit non-perturbative vacuum massive solutions were constructed in the case of gravitational Higgs mechanism in Minkowski space.\footnote{\, That is, Minkowski space being one of the background solutions.} There appears to be no ``instability" or catastrophic collapse of the background. Instead, non-perturbatively there simply exist vacuum solutions other than the Minkowski background. These vacuum solutions are oscillatory with no evident pathologies or cause for alarm. That is, despite the fact that perturbatively there is a fake ghost, non-perturbatively the theory does not appear to be ``sick" or inconsistent. In fact, it has a rich structure of apparently well-behaved (non-static) vacuum solutions.\footnote{\, Whether there are any tunneling effects is a separate and interesting topic for investigation.}

{}Motivated by these results and also by the observation of \cite{ZK.vDVZ}, in this note we discuss non-perturbative dynamics of massive gravity in de Sitter space via gravitational Higgs mechanism. We keep $\beta$ arbitrary. For $\beta=1$, perturbatively there is the Higuchi bound \cite{Higuchi}, according to which the helicity-0 graviton mode becomes null at $M^2 = 2H^2$ and turns into a ghost for $M^2 < 2H^2$ (where $H$ is the Hubble parameter in de Sitter space), which is a form of vDVZ discontinuity \cite{vDV, Zak}. The appearance of the null state at $M^2 = 2H^2$ is due to enhanced local symmetry \cite{DN,Higuchi,DW}. However, as was argued in \cite{dS}, non-perturbatively there is no enhanced symmetry or null state for any value of the graviton mass and these are merely artifacts of linearization as is the perturbative ghost. In this paper we first consider the linearized theory for general $\beta$ and argue that for $\beta\neq 1$ there is no enhanced symmetry or null state already at the linearized level. So, in this regard, these are artifacts of not only linearization but also assuming the Fierz-Pauli mass term.

{}We then discuss full non-perturbative equations of motion for general $\beta$ and $M^2/H^2$. For general $\beta$ and dimension $D$ the analog of the perturbative Higuchi bound is
\begin{equation}\label{gen.Higuchi}
 M_*^2 = {(D-1)(D-2)\over{D\beta-1}}~H^2~.
\end{equation}
However, we emphasize that non-perturbatively there is no enhanced local symmetry or null state at $M^2 = M_*^2$, and there is no ghost at $M^2 < M_*^2$. Instead, the significance of the ``special point" $M^2 = M_*^2$ is that it delineates the types of vacuum solutions other than de Sitter that exist. Thus, we argue that there exist vacuum solutions which asymptotically are de Sitter for both ${\widetilde \tau}\rightarrow \pm\infty$, where ${\widetilde \tau}$ is the ``proper time", and have the property that the space (starting from de Sitter at ${\widetilde \tau}\rightarrow -\infty$) first contracts, this contraction slows down, and then reverses into expansion (ending with de Sitter at ${\widetilde \tau}\rightarrow +\infty$), so there is an epoch where the space appears to be (nearly) flat, even though the vacuum energy density is non-vanishing.\footnote{\, An interesting feature of these solutions is that the scalar curvature at ${\widetilde \tau}\rightarrow +\infty$ is about two orders of magnitude smaller than at ${\widetilde \tau}\rightarrow -\infty$, even though there is no small parameter in the theory -- see Subsection 4.1 for details.} Such solutions exist for all $M^2 \leq M_c^2$, where $M_c$ is the ``critical" graviton mass. It is convenient to parameterize $M^2$ via $1-\alpha\equiv M_*^2/M^2$, so $\alpha \in (-\infty, 1)$, and $\alpha = 0$ corresponds to $M^2 = M_*^2$. We derive an exact formula for the value $\alpha_c$ corresponding to $M_c^2$ for arbitrary $\beta$ and $D$ (see Section 5, Eq. (\ref{alpha.crit.gen}) for details):
\begin{equation}
 \alpha_c\equiv {{D-1}\over{D+{{3+\beta}\over{1-\beta}}}}~,
\end{equation}
so the aforementioned vacuum solutions do not exist for $\alpha > \alpha_c$, or for $M^2 > M_c^2$, where
\begin{equation}
 M_c^2 = {M_*^2\over{1-\alpha_c}} = {(D-1)(D-2)\over{4(D\beta-1)}}\left(D(1-\beta)+ 3+\beta\right)H^2~.
\end{equation}
For $\beta=1$ (the Fierz-Pauli case) we have $M_c^2 = (D-2)H^2 = M_*^2$~,
so there are no such asymptotic solutions for $M^2 > M_*^2$ in this case, while for $M^2\leq M_*^2$ we do have such solutions. And non-perturbatively this is the meaning of the Higuchi bound for $\beta=1$. For $\beta\neq 1$ it is $M_c^2$, not $M_*^2$, that carries this meaning, albeit $M_*^2$ defined in (\ref{gen.Higuchi}) does serve as a more subtle ``demarkation line" for more detailed properties of these vacuum solutions (see Section 4 for details).

{}In fact, for $\beta = 1/2$ and $\alpha = 0$ ({\em i.e.}, at the ``special point" $M^2=M_*^2$) we solve the full non-perturbative equations of motion in closed form and explicitly construct the solution described above (see Eq. (\ref{Q}) and the subsequent discussion for details). (We have also constructed it numerically as an additional check, see Fig.1.) Just as in the Minkowski case \cite{ZK.massive}, we find no ``instability" or catastrophic collapse of the background. Instead, non-perturbatively there simply exist vacuum solutions other than de Sitter. An interesting feature of these solutions is the initial contraction and the subsequent expansion, with an epoch where the space-time appears to be (nearly) flat. And this is irrespective of the value of the cosmological constant. Such solutions may be interesting in the context of the cosmological constant problem. {\em E.g.}, one could imagine that we live in such a universe and the ``observed value" of the cosmological constant ({\em i.e.}, the accelerated rate of expansion) is small because we are just past the turning point and starting to expand. Another possible application could be in the context of inflation and the Big Bang. In particular, one could imagine a scenario where the space contracts to a small (Planckian) size and then expands again, providing the ``starting point" for the Universe. Just speculations\dots

{}To summarize, despite the fact that perturbatively there is a fake ghost, non-perturbatively the theory does not appear to be ``sick" or inconsistent.
In fact, it has a rich structure of (possibly interesting) vacuum solutions. The remainder of this note is organized as follows. In Section 2 we very briefly review gravitational Higgs mechanism in curved backgrounds. In Section 3 we discuss linearized massive gravity in de Sitter space. In Section 4 we discuss non-perturbative dynamics in the $\beta=1/2$ case and give the explicit exact solution for $\alpha=0$. In Section 5 we derive the formulas for $\alpha_c$ and $M^2_c$ for general $\beta$.

\section{Gravitational Higgs Mechanism}

{}In this section we very briefly review gravitational Higgs mechanism in curved backgrounds -- de Sitter was discussed in \cite{dS}, and the general case in \cite{ZK.massive}. We start with gravity in $D$ dimensions
\begin{equation}\label{action.0}
 S_G \equiv M_P^{D-2}\int d^Dx \sqrt{-G}\left[ R - {\widetilde \Lambda}\right]~,
\end{equation}
where ${\widetilde \Lambda}$ is the cosmological constant. Let ${\widetilde G}_{MN}$ be a background solution to the equations of motion corresponding to (\ref{action.0}). The background metric ${\widetilde G}_{MN}$ generally is a function of the coordinates $x^S$: ${\widetilde G}_{MN} = {\widetilde G}_{MN}(x^0,\dots, x^{D - 1})$.

{}Let us now introduce $D$ scalars $\phi^A$. Let us normalize them such that they have dimension of length. Let us define a metric $Z_{AB}$ for the scalars as follows:
\begin{equation}
 Z_{AB}(\phi^0,\dots,\phi^{D-1}) \equiv {\delta_A}^M {\delta_B}^N {\widetilde G}_{MN}(\phi^0,\dots, \phi^{D - 1})~.
\end{equation}
Next, consider the induced metric for the scalar sector:
\begin{equation}
 Y_{MN} = Z_{AB} \nabla_M\phi^A \nabla_N\phi^B~.
\end{equation}
The following action, albeit not most general,\footnote{\label{foot}One can consider a
more general setup with the scalar action constructed not just from $Y$,
but from $Y_{MN}$, $G_{MN}$ and $\epsilon_{M_0\dots M_{D-1}}$.} will suffice for our purposes here:
\begin{equation}
 S_Y = M_P^{D-2}\int d^Dx \sqrt{-G}\left[ R - \mu^2 V(Y) \right]~,
 \label{actionphiY}
\end{equation}
where {\em a priori} the ``potential" $V(Y)$ is a generic function of $Y\equiv Y_{MN}G^{MN}$, and $\mu$ is a mass parameter, while ${\widetilde \Lambda}$ is subsumed in the definition of $V(Y)$.

{}The equations of motion have the following solutions
\begin{eqnarray}\label{solphiY}
 &&\phi^A = {\delta^A}_M~x^M~,\\
 \label{solGY}
 &&G_{MN} = {\widetilde G}_{MN}~,
\end{eqnarray}
with
\begin{equation}\label{cosm.const}
 {\widetilde \Lambda} = \mu^2\left[V(D) - 2 V^\prime(D)\right]~.
\end{equation}
The scalar vacuum expectation values break diffeomorphism spontaneously, and the equations of motion are invariant under the full diffeomorphism invariance. We can therefore set the scalars to their background values, which leaves us with pure gravity:
\begin{equation}
 S_{MG} = M_P^{D-2}\int d^Dx \sqrt{-G}\left[ R - \mu^2 V(X) \right]~,
\end{equation}
where $X\equiv G^{MN}{\widetilde G}_{MN}$. The equations of motion read
\begin{equation}\label{R.eom}
 R_{MN} = \mu^2 \left[V^\prime(X)~{\widetilde G}_{MN} +  {{V(X) - X~V^\prime(X)}\over {D-2}}~G_{MN}\right]~,
\end{equation}
with the Bianchi identity
\begin{equation}\label{phi.eom}
 \partial_M\left[\sqrt{-G}V^\prime(X) G^{MN}{\widetilde G}_{NS}\right] - {1\over 2}\sqrt{-G}V^\prime(X) G^{MN}\partial_S {\widetilde G}_{MN} = 0~,
\end{equation}
which is equivalent to the gauge-fixed equations of motion for the scalars.

{}In the linearized approximation in (\ref{R.eom}) we have the mass term corresponding to (\ref{mass.term}) with
 \begin{eqnarray}\label{m}
 &&M^2\equiv 2\mu^2 V^\prime(D)~,\\
 &&\beta \equiv {1\over 2} - {V^{\prime\prime}(D) \over {V^\prime(D)}}~.\label{beta}
\end{eqnarray}
We have $\beta = 1$ for potentials $V$ with $V^\prime(D) = -2V^{\prime\prime}(D)$. For a linear potential $V(X) = a + X$, we have the model of \cite{thooft} with $\beta = 1/2$ (and for $a = -(D-2)$ we have the Minkowski background). {\em E.g.}, for quadratic potentials $V = a + X + \lambda X^2$ with $\lambda\neq 0$ we can have other values of $\beta$, including $\beta = 1$ for $\lambda = -1/2(D+2)$.

\section{Massive Gravity in de Sitter}

{}In this section we study linearized equations of motion for the de Sitter background metric ${\widetilde G}_{MN}\equiv (Ht)^{-2}\eta_{MN}$, where $t\equiv x^0$, $H$ is the constant Hubble parameter, $H^2\equiv{\widetilde\Lambda}/(D-1)(D-2)$, and $\eta_{MN}$ is the Minkowski metric. Let us keep $\beta$ in (\ref{beta}) arbitrary for now. Let $G_{MN} \equiv {\widetilde G}_{MN} + h_{MN}$ and $h \equiv {\widetilde G}^{MN} h_{MN}$. Then the linearized equations of motion (\ref{R.eom}) read:
\begin{eqnarray}\label{lin.eq}
 &&S_{MN} = 0~,\\
 &&S_{MN}\equiv \Box ~h_{MN} + \nabla_M\nabla_N h - 2\nabla_{(M}\nabla^S h_{N)S} -
 {\widetilde G}_{MN}\left[\Box ~h - \nabla^S\nabla^R h_{SR}\right] - \nonumber\\
 &&\,\,\,\,\,\,\,H^2 \left[2 h_{MN} + (D-3) {\widetilde G}_{MN} h \right] - M^2 \left[h_{MN} - \beta{\widetilde G}_{MN} h\right]~,\label{S_MN}
\end{eqnarray}
where $\nabla_M$ is the covariant derivative in the de Sitter background metric ${\widetilde G}_{MN}$, and $\Box \equiv {\widetilde G}^{MN} \nabla_M \nabla_N$. Furthermore, the Bianchi identity (\ref{phi.eom}) reduces to
\begin{equation}\label{lin.con}
 \nabla^N h_{MN} - \beta \nabla_M h = 0~.
\end{equation}
Note that (\ref{lin.con}) follows from (\ref{lin.eq}).

\subsection{Enhanced Local Symmetry at $\beta = 1$}

{}For the Fierz-Pauli mass terms ($\beta = 1$) at a special value of the ratio ${\widetilde \Lambda}^2/ M^2$ the linearized equations of motion (\ref{lin.eq}) are invariant under the following infinitesimal local transformations:
\begin{equation}\label{special}
 \delta h_{MN} = (\nabla_M\nabla_N +  H^2 {\widetilde G}_{MN})\chi~.
\end{equation}
This local symmetry is present when
\begin{equation}\label{M-H}
 M^2 = (D-2) H^2~.
\end{equation}
The following identities are useful in deriving this result:
\begin{eqnarray}\label{identity.1}
 &&\left(\Box~\nabla_M - \nabla_M~\Box\right)\chi = (D-1) H^2~\nabla_M \chi~,\\
 &&\left(\Box~\nabla_M\nabla_N - \nabla_M\nabla_N~\Box\right)\chi = 2 H^2\left(D~\nabla_M\nabla_N - {\widetilde G}_{MN}~\Box\right)\chi~.
\end{eqnarray}
In $D=4$ the presence of the symmetry (\ref{special}) at the point (\ref{M-H}) was discussed in \cite{DN}.

{}The presence of this additional local symmetry in the linearized theory implies that at the point (\ref{M-H}) the graviton has $(D+1)(D-2)/2 - 1$ propagating degrees of freedom (as the helicity-0 graviton mode has null norm), one fewer than at generic points in the parameter space. Furthermore, the linearized theory is non-unitary for $M^2 < (D-2) H^2$ (as the helicity-0 graviton mode has negative norm), while for $M^2 > (D-2) H^2$ all $(D+1)(D-2)/2$ graviton modes are propagating and have positive norm \cite{DN, Higuchi, DW}. Also, in the linearized theory at $M^2 = (D-2) H^2$ the graviton $h_{MN}$ can only couple to traceless conserved energy-momentum tensor $T_{MN}$ as the coupling to the trace part of $T_{MN}$ is inconsistent with the symmetry (\ref{special}). For a recent discussion, see \cite{GI, GIS}.

{}However, as was argued in \cite{dS}, the appearance of the enhanced local symmetry at $M^2 = (D-2) H^2$ is an artifact of linearization. There is no enhanced local symmetry in the full nonlinear theory, nor is there a null state or a ghost, and the full nonlinear theory unitary is for all values of $M^2/H^2$ with no vDVZ discontinuity. The purpose of our exercise here is to argue that the enhanced local symmetry (and all its consequences) is not only an artifact of linearization but also of requiring the Fierz-Pauli term. This is reminiscent to what transpires in static radially symmetric solutions with ${\widetilde \Lambda} = 0$ -- as was argued in \cite{ZK.vDVZ}, there is no vDVZ discontinuity in the perturbative asymptotic solutions except for $\beta = 1$.

\subsection{General $\beta$}

{}It is not difficult to see that the symmetry
\begin{equation}
 \delta h_{MN} = (\nabla_M\nabla_N +  \gamma H^2 {\widetilde G}_{MN})\chi~,
\end{equation}
where $\gamma$ is an arbitrary dimensionless constant, is not a symmetry of the linearized equations of motion for $\beta\neq 1$. Indeed, (\ref{lin.con}) is not invariant under these transformations unless $\beta = 1$. It then follows that there cannot be any null state for $\beta\neq 1$. Without a null state we do not expect a transition between the regime without a ghost to a regime with a ghost either. {\em I.e.}, what transpires at $M^2 = (D-2)H^2$ in the $\beta = 1$ case is an artifact of $\beta = 1$ and does not happen at other values of $\beta$.

{}We can see this explicitly by studying the linearized equations of motion. The simplest way to proceed is to focus on the relevant graviton degrees of freedom, namely, the conformal mode $\omega$ and the helicity-0 mode $\rho$. These two modes are related via (\ref{lin.con}). Let $h_{MN} = {\widetilde G}_{MN}\omega + \nabla_M\nabla_N\rho$. Then (\ref{lin.con}) gives
\begin{equation}\label{omega-rho}
 (D\beta - 1)\omega = (D-1)H^2\rho + (1-\beta)\Box\rho~.
\end{equation}
Note that only for $\beta = 1$ is the relationship between $\omega$ and $\rho$ algebraic. It is precisely this feature that leads to ``funky" behavior for $\beta = 1$.

{}To see this more explicitly, let us look at the full equations of motion. The equation of motion for the trace $h$ reads:
\begin{equation}\label{h}
 -(D-2)(1-\beta)\Box h  + \left[(D\beta - 1)M^2 - (D-1)(D-2)H^2\right] h = 0~.
\end{equation}
Again, for $\beta = 1$ this equation is algebraic and generally implies that $h = 0$. At the special point with $\beta = 1$ and $M^2 = (D-2)H^2$, this equation reduces to an identity, which is a manifestation of the enhanced local symmetry at this point. Furthermore, if we included a matter source, the r.h.s. of (\ref{h}) would be proportional to the trace of the conserved energy-momentum tensor, which therefore must vanish for $\beta = 1$ and $M^2 = (D-2)H^2$. However, for $\beta \neq 1$ nothing of the kind transpires as the $\Box h$ term is present regardless of the value of $M^2/H^2$.

{}Let us take this a step further and analyze $S_{MN}$ in (\ref{S_MN}) for the modes $\omega$ and $\rho$. A little algebra gives
\begin{equation}
 S_{MN} = (D-2)\left[\nabla_M\nabla_N~{\overline\omega} - {\widetilde G}_{MN}\Box~{\overline \omega}\right] - (D-1)(D-2)H^2{\widetilde G}_{MN}~{\overline \omega}~,
\end{equation}
where we have used (\ref{identity.1}) and ({\ref{omega-rho}), and
\begin{equation}\label{omega.bar}
 {\overline \omega} \equiv \omega - {M^2\over{D-2}}~\rho = \left[{{D-1}\over{D\beta - 1}}~H^2 - {M^2\over{D-2}}\right] \rho + {{1-\beta}\over{D\beta - 1}}~
 \Box\rho~.
\end{equation}
Note that when $\beta = 1$, we have ${\overline \omega} = \left[H^2 - M^2/(D-2)\right]\rho$, so $\rho$ is null at $M^2 = (D-2)H^2$, is a ghost at $M^2 < (D-2)H^2$, and is a positive-norm state for $M^2 > (D-2)H^2$. However, for $\beta \neq 1$ there is no null state. Instead, we have a higher-derivative equation of motion for $\rho$. We will not analyze this {\em linearized} higher-derivative equation of motion from the viewpoint of stability -- as was argued in \cite{dS}, stability needs to be addressed non-perturbatively (and was discussed in detail in \cite{dS}) as the linearization artificially introduces a ghost,\footnote{\,\,\, In fact, as was argued in \cite{Unitarity}, the intrinsically perturbative parametrization $h_{MN} = {\widetilde G}_{MN}\omega + \nabla_M\nabla_N\rho$ is inadequate for addressing stability.} including for $\beta\neq 1$. And we will discuss non-perturbative solutions momentarily. Our key observation here is that the enhanced local symmetry at the special point and its consequences are merely an artifact of requiring the Fierz-Pauli mass term.

\section{Non-perturbative Dynamics}

{}Looking at (\ref{h}) or (\ref{omega.bar}), one might get an uneasy feeling that even for $\beta\neq 1$ something funny might transpire as the graviton mass crosses the ``special point" defined as
\begin{equation}
 M^2 = M_*^2 \equiv{(D-1)(D-2)H^2\over{D\beta- 1}} = {{\widetilde \Lambda}\over {D\beta- 1}}~.
\end{equation}
{\em E.g.}, one may even worry about a potential tachyonic instability. As mentioned above, stability must be analyzed non-perturbatively, which was done in \cite{dS} by studying the non-perturbative Hamiltonian, which was argued to be bounded from below, hence no ghost or instability. Here we will not repeat the analysis of \cite{dS}, but instead complement it by studying solutions to full non-perturbative equations of motion, similarly to the Minkowski case in \cite{ZK.massive}.

{}The equations of motion (\ref{R.eom}) are highly nonlinear. However, we can get a handle on them by considering linear potential $V = a + X$ \cite{thooft}. In this case we have ${\widetilde \Lambda} = (a + D - 2)\mu^2$, $M^2 = 2\mu^2$, $\beta = 1/2$ and $M_*^2 / M^2 = 1 - \alpha$, where $\alpha\equiv -a / (D-2)$, so the ``special point" corresponds to $\alpha = 0$.

{}The equations of motion (\ref{R.eom}) read
\begin{equation}\label{Ricci}
 R_{MN} = \mu^2\left[{\widetilde G}_{MN} - \alpha G_{MN}\right].
\end{equation}
The naively ``troublesome" case is when $\alpha$ is negative.

\subsection{``Special Point": $\alpha = 0$ (for $\beta = 1/2$)}

{}Let us start with studying non-perturbative solutions at the ``special point" $\alpha = 0$. We will look for the solutions of the form
\begin{equation}\label{ansatz}
 G_{MN} = \mbox{diag}\left(\exp(2B)\eta_{00}, \exp(2A)\eta_{ii}\right).
\end{equation}
Then we have $R_{i0} = 0$ and (prime denotes derivative w.r.t. $t$):
\begin{eqnarray}
 &&R_{00} = -(D-1)\left[A^{\prime\prime} - A^\prime B^\prime + \left(A^\prime\right)^2\right],\\
 &&R_{ij} = \left[A^{\prime\prime} - A^\prime B^\prime + (D-1)\left(A^\prime\right)^2\right]\exp(2A - 2B)~\delta_{ij}~,
\end{eqnarray}
and the $(00)$ and the $(ij)$ equations of motion read:
\begin{eqnarray}
 &&A^{\prime\prime} - A^\prime B^\prime + \left(A^\prime\right)^2 = t^{-2}~,\\
 &&A^{\prime\prime} - A^\prime B^\prime + (D-1)\left(A^\prime\right)^2 = (D-1)\exp(2B-2A)~t^{-2}~,
\end{eqnarray}
where we have taken into account that at $\alpha = 0$ we have $\mu^2 = (D-1)H^2$. The Bianchi identity (\ref{phi.eom}) reduces to
\begin{equation}
 t^{-1}\left[(D-1)\exp(2B-2A) - 1\right] + (D-1)A^\prime - B^\prime = 0~,
\end{equation}
which together with the equations of motions allows to eliminate $B^\prime$:
\begin{equation}
 B^\prime = (D-2)t\left(A^\prime\right)^2 + (D-1)A^\prime~.
\end{equation}
The $(00)$ equation then reduces to the following first-order nonlinear equation:
\begin{equation}\label{Qtau}
 Q_\tau = \left[1 + Q\right]\left[1 + (D-2)Q^2\right]~,
\end{equation}
where $Q\equiv tA^\prime = A_\tau\equiv \partial_\tau A$, and $Q_\tau \equiv\partial_\tau Q$, where $\partial_\tau\equiv t\partial_t$. Note that the background solution $G_{MN} = {\widetilde G}_{MN}$ corresponds to the $Q=-1$ solution of (\ref{Qtau}). However, we have another vacuum solution given by
\begin{eqnarray}
 &&\ln\left|1 + Q\right| - {1\over 2}\ln\left((D-2)^{-1} + Q^2\right) + \sqrt{D-2}\tan^{-1}\left(\sqrt{D-2}Q\right) = \nonumber\\
 &&\,\,\,\,\,\,\,(D-1)\left(\tau - \tau_0\right)~,\label{Q}
\end{eqnarray}
where $\tau_0$ is an integration constant. In fact, there are two inequivalent solutions, which we will refer to as $Q^{(\pm)}$, with the following properties.
The $Q^{(\pm)}$ solutions are defined for $\tau\in(-\infty,\tau_*^{(\pm)})$, where $\tau_*^{(\pm)}\equiv \tau_0^{(\pm)}\pm\sqrt{D-2}\pi/2(D-1)$, and we have $Q^{(+)} > -1$, $Q^{(-)} < -1$, at $\tau\rightarrow-\infty$ we have $Q^{(\pm)} \rightarrow -1\pm$, $Q^{(+)}$ monotonically increases from $-1$ to $+\infty$, and $Q^{(-)}$ monotonically deceases from $-1$ to $-\infty$. As we will see in a moment, the singularities at $\tau_*^{(\pm)}$ are not true singularities but mere coordinate singularities with finite ``proper time". To obtain a geodesically complete space, we must continue each solution through this coordinate singularity. To do this, let us study the behavior of the metric at $\tau\rightarrow \tau_*^{(\pm)}-$. We have $\left(Q^{(\pm)}\right)^2 \sim 1/2(D-2)(\tau_*^{(\pm)} - \tau)$ and $Q^{(\pm)}\sim \pm 1/\sqrt{2(D-2)(\tau_*^{(\pm)} - \tau)}$, which implies that $A^{(\pm)}\sim A^{(\pm)}_* \mp \sqrt{2(\tau_*^{(\pm)} - \tau)/(D-2)}$, where $A^{(\pm)}_*$ are constants, so $A^{(\pm)}$ are finite at $\tau\rightarrow \tau^{(\pm)}_*-$, albeit their derivatives w.r.t. $\tau$ are not. Furthermore, we have $B^{(\pm)}\sim -\ln\left(\tau^{(\pm)}_* - \tau\right)/2$, so $B^{(\pm)}$ are singular. However, this is just a coordinate singularity due to the choice of the time coordinate $\tau$. The Ricci tensor is finite and so is the scalar curvature.\footnote{\, Before fixing the gauge by setting the scalars to their background values via (\ref{solphiY}), we have additional gauge invariants $G^{MN}\partial_M \phi^A \partial_N \phi^B$, which upon gauge fixing reduce to $G^{MN}$. Since $A$ is finite and $B$ goes to $+\infty$ at $\tau\rightarrow \tau^{(\pm)}_*-$, these additional gauge invariants ({\em i.e.}, the components of the inverse metric) are also finite.} Furthermore, this coordinate singularity is at finite ``proper time". Thus, first note that $\tau = \ln\left(\nu^{(\pm)} t\right)$, where $\nu^{(\pm)}$ are integration constants for the $Q^{(\pm)}$ solutions. Let $t_*^{(\pm)}$ correspond to $\tau_*^{(\pm)}$: $t_*^{(\pm)} \equiv \exp(\tau_*^{(\pm)})/\nu^{(\pm)}$. Let us further define the ``proper time" $d{\widetilde \tau} \equiv\exp\left(B^{(\pm)}\right)dt$. Since, $dt = t d\tau$, as $\tau\rightarrow \tau_*^{(\pm)}-$, we have $d{\widetilde \tau} \sim t_*^{(\pm)} \exp\left(B^{(\pm)}\right)d\tau \sim t_*^{(\pm)} d\tau/\sqrt{\tau_*^{(\pm)} - \tau}$, and ${\widetilde\tau} \sim {\widetilde\tau}_*^{(\pm)} - 2t_*^{(\pm)} \sqrt{\tau_*^{(\pm)} - \tau}$, where ${\widetilde\tau}_*^{(\pm)}$ are constants corresponding to $\tau_*^{(\pm)}$, {\em i.e.}, these values of the ``proper time" are finite. Also, note that $A^{(\pm)}_{\widetilde\tau} \sim \pm 1/\sqrt{2(D-2)}t_*^{(\pm)}$, so the derivative of $A$ w.r.t. the ``proper time" is finite. Furthermore, note that at $\tau\rightarrow-\infty$ we have $B^{(\pm)} \rightarrow -\tau$, and $d{\widetilde\tau}\sim d\tau/\nu^{(\pm)}$, so even though $\tau\rightarrow-\infty$, we can have ${\widetilde\tau}\rightarrow+\infty$ or ${\widetilde\tau}\rightarrow-\infty$ depending on the sign of $\nu^{(\pm)}$.

{}The metric in the ``proper time" (or flat slicing) coordinates is given by
\begin{equation}\label{flat.slicing}
 ds^2 \sim -d{\widetilde \tau}^2 + \exp(2A)~\delta_{ij}dx^idx^j~,
\end{equation}
which is finite. Now we can sew the $Q^{(\pm)}$ solutions together to obtain a geodesically complete space. Thus, consider the following solution. Let $A^{(+)}_* = A^{(-)}_* \equiv A_*$. $t^{(+)}_* = -t^{(-)}_* \equiv t_*$, $\nu^{(+)} = -\nu^{(-)} \equiv \nu$, $\tau^{(+)}_* = \tau^{(-)}_* \equiv \tau_*$ (which implies that $\tau_0\equiv \tau^{(+)}_0 = \tau^{(-)}_0 -\sqrt{D-2}\pi/(D-1)$). Furthermore, let ${\widetilde \tau}^{(+)}_* = {\widetilde \tau}^{(-)}_* \equiv {\widetilde \tau}_*$. Then we have a smooth solution defined for ${\widetilde \tau} \in {\bf R}$ such that $Q\equiv A_\tau$ is given by $Q = Q^{(\pm)}$ for ${\widetilde \tau} \in {\bf R}^{\mp}_{\widetilde \tau}$ when $\nu > 0$, and $Q = Q^{(\pm)}$ for ${\widetilde \tau} \in {\bf R}^{\pm}_{\widetilde \tau}$ when $\nu < 0$, where ${\bf R}^-_x \equiv (-\infty, x]$ and ${\bf R}^+_x \equiv [x, +\infty)$. Note that $A$ is continuous and $A_{\widetilde \tau}$ is also continuous (even though $A_\tau$ is not). The $\nu > 0$ and $\nu < 0$ solutions are mirror to each other under ${\widetilde \tau} \rightarrow -{\widetilde \tau}$. For definiteness, without loss of generality, we will therefore assume $\nu > 0$.
\begin{figure}
\centerline{\epsfxsize 4.truein \epsfysize 4.truein\epsfbox{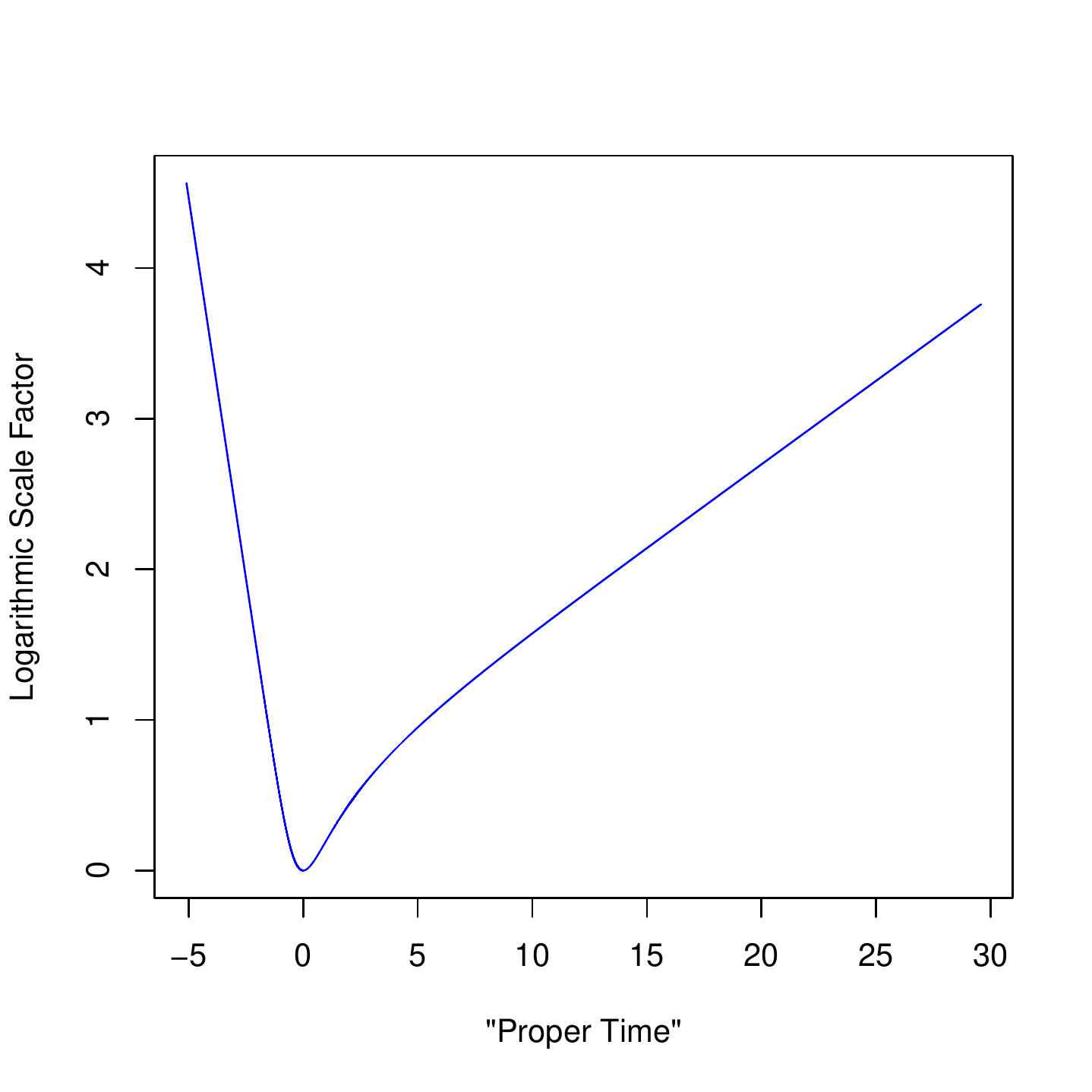}}
\caption{Exact solution ($D=4$, $\nu=1$) of Subsection 4.1. $x$-axis: ``proper time" ${\widetilde\tau}$; $y$-axis: $A$ in (\ref{flat.slicing}). (We have set ${\widetilde\tau}_1$ and $A({\widetilde\tau}_1)$ to 0 in (\ref{flat.slicing}) -- see Subsection 4.1.)}
\end{figure}

{}In this solution at ${\widetilde \tau} \rightarrow -\infty$ we have de Sitter space. As ${\widetilde \tau}$ increases, the space contracts, but this contraction slows down and stops at ${\widetilde \tau} = {\widetilde \tau}_1$, where ${\widetilde \tau}_1$ corresponds to $\tau_1 \equiv \tau_0 + \ln(D-2)/2(D-1) < \tau_* \equiv \tau_0 + \sqrt{D-2}\pi/2(D-1)$, and for ${\widetilde \tau} > {\widetilde \tau}_1$ the space starts to expand, seamlessly goes through the ``sewing" point ${\widetilde\tau} = {\widetilde\tau}_*$ and continues to expand asymptotically approaching de Sitter at ${\widetilde \tau} \rightarrow +\infty$. The interesting feature of this solution is that there is an epoch where the space appears to be (nearly) flat, which is near ${\widetilde \tau} = {\widetilde \tau}_1$. And this occurs regardless of the value of the cosmological constant ${\widetilde \Lambda}$. In fact, this solution is a ``no-scale" solution despite the fact that there is a mass scale in the theory. In particular, if we rescale $\nu\rightarrow\lambda\nu$, where $\lambda$ is a constant, the solution is the same as before with the rescaling ${\widetilde \tau}\rightarrow {\widetilde \tau} /\lambda$. Fig.1 shows this solution with $D=4$, $\nu = 1$, ${\widetilde \tau}_1= 0$ and $A({\widetilde \tau}_1) = 0$. Fig.2 shows $\ln(R/\mu^2)$. Even though there is no small parameter, the scalar curvature $R$ drops about two orders of magnitude from $R({\widetilde\tau}\rightarrow-\infty) \approx 10.2~\mu^2$ to $R({\widetilde\tau}\rightarrow+\infty)\approx 0.123~\mu^2$.
\begin{figure}
\centerline{\epsfxsize 4.truein \epsfysize 4.truein\epsfbox{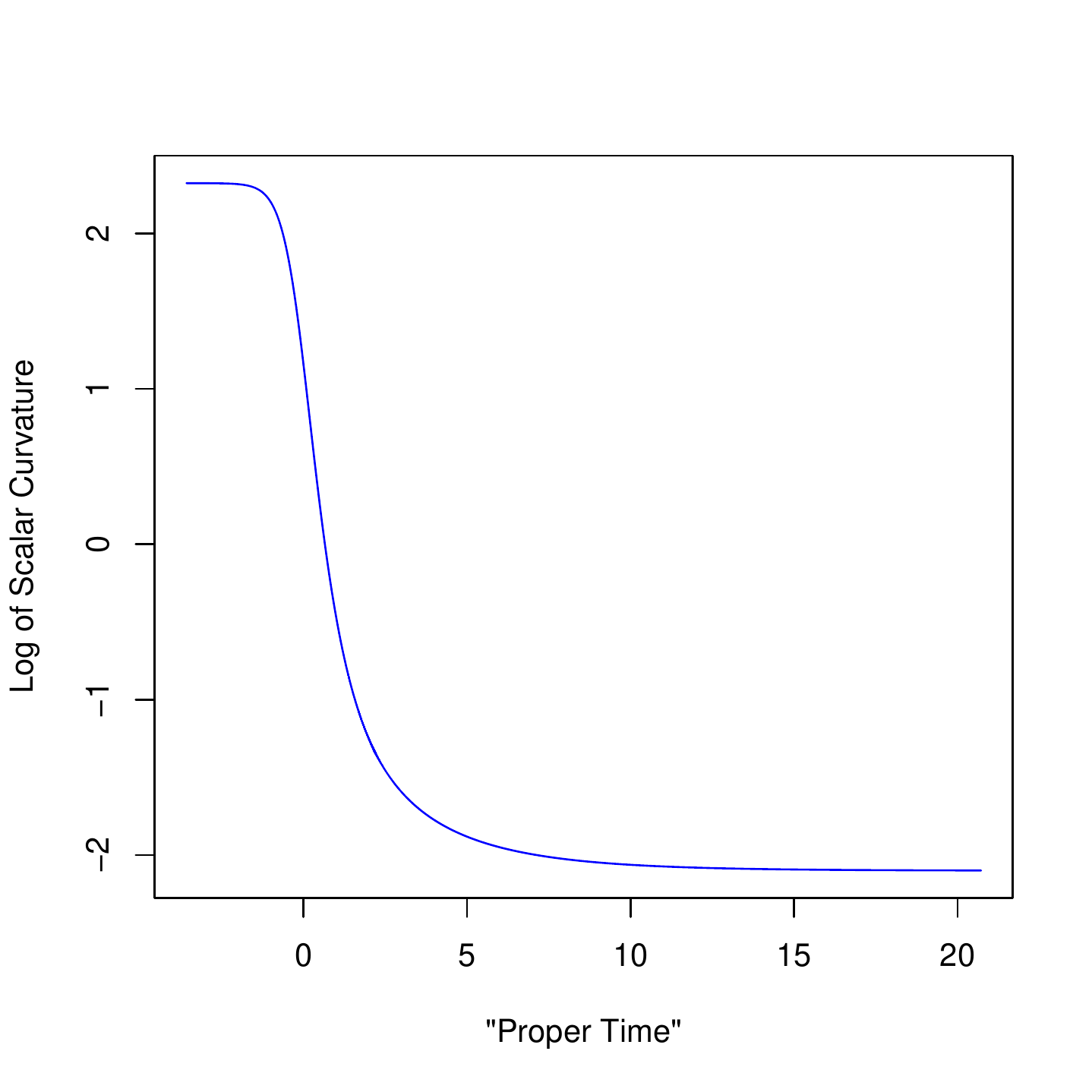}}
\caption{Scalar curvature via (\ref{Ricci}) in the Fig.1 solution. $x$-axis: ${\widetilde\tau}$; $y$-axis: $\ln(R/\mu^2)$.}
\end{figure}

\subsection{General $\alpha$ (for $\beta = 1/2$)}

{}In the case of general $\alpha$, where $\alpha \in(-\infty, 1)$, the above equations of motion for $Q$ and $B$ are more complicated:
\begin{eqnarray}\label{Q1}
 &&Q_\tau = {(D-1)(Q+1)\over(1-\alpha)(D-2)} + \left[Q - {1\over(1-\alpha)(D-2)}\right]{\widetilde B}_\tau~,\\
 &&{\widetilde B}_\tau = (1-\alpha)(D-2)Q^2 + (D-1)Q + \alpha(D-2)e^{2{\widetilde B}} + 1~,\label{B1}
\end{eqnarray}
where ${\widetilde B} \equiv B + \ln(Ht)$. It will also be useful to rewrite (\ref{Q1}) and (\ref{B1}) as follows:
\begin{eqnarray}\label{Q2}
 &&Q_\tau = (Q+1)\left[(1-\alpha)(D-2)Q^2 + \alpha(D-2)Q + 1\right]+\nonumber\\
 &&\,\,\,\,\,\,\,\alpha(D-2)\left[Q - {1\over(1-\alpha)(D-2)}\right]\left[e^{2{\widetilde B}} - 1\right].\\
 &&{\widetilde B}_\tau = (Q+1)\left[(1-\alpha)(D-2)Q + \alpha(D-2) + 1\right] +\nonumber\\
 &&\,\,\,\,\,\,\,\alpha(D-2)\left[e^{2{\widetilde B}}-1\right].\label{B2}
\end{eqnarray}
So we have a system of two first order nonlinear differential equations, which is difficult to solve in closed form for general $\alpha$. (Note that for $\alpha =0$ the exponential term vanishes and the system reduces to a first order nonlinear differential equation for $Q$, which we solved above.) Happily, we do not need to solve this system in closed form to understand if there is a ghost or tachyonic instability for $\alpha < 0$ (or $\alpha > 0$ for that matter).

{}First, note that irrespective of the value of $\alpha$ we have the de Sitter solution where $Q\equiv -1$ and ${\widetilde B}\equiv 0$. The question is whether there are any other solutions as in the $\alpha = 0$ case. Asymptotically, at $|\tau|\rightarrow \infty$, {\em a priori} we can have $Q\rightarrow\mbox{const.}$ or $|Q|\rightarrow \infty$. Let us show that asymptotically we cannot have $|Q|\rightarrow \infty$. From (\ref{Q1}) we have ${\widetilde B} \sim \ln|Q| - (D-1)(\tau-\tau_0)/(1-\alpha)(D-2)$ in this case, where $\tau_0$ is an integration constant. Then (\ref{B1}) implies that $Q_\tau\approx(D-2) Q^3 \left[1 - \alpha + \alpha\exp\left(-2(D-1)(\tau-\tau_0)/(1-\alpha)(D-2)\right)\right]$. This, in turn, implies that $|Q|\rightarrow\infty$ must occur at finite $\tau\rightarrow \tau_*$. Indeed, let us first assume that $|Q|\rightarrow\infty$ occurs at $\tau\rightarrow+\infty$. Then we would have $Q_\tau\approx
(1 - \alpha)(D-2)Q^3$, whose solution behaves as $1/Q^2\sim 2(1-\alpha)(D-2)(\tau_*-\tau)$, where $\tau_*$ is an integration constant, so our assumption that $|Q|\rightarrow\infty$ occurs at $\tau\rightarrow+\infty$ does not hold. Next, let us assume that $|Q|\rightarrow\infty$ occurs at $\tau\rightarrow-\infty$. Then we would have $Q_\tau\approx (D-2)\alpha Q^3 \exp\left(-2(D-1)(\tau-\tau_0)/(1-\alpha)(D-2)\right)$, whose solution behaves as $1/Q^2\sim \alpha(1-\alpha)(D-2)^2\exp\left(-2(D-1)(\tau-\tau_0)/(1-\alpha)(D-2)\right)/(D-1)$, so our assumption that $|Q|\rightarrow\infty$ occurs at $\tau\rightarrow-\infty$ also cannot be correct. Since $|Q|\rightarrow\infty$ can only occur at finite $\tau\rightarrow \tau_*$, we have
$1/Q^2 \sim 2(D-2)\zeta(\tau_*-\tau)$, where $\zeta\equiv 1-\alpha+ \alpha\exp\left(-2(D-1)(\tau_*-\tau_0)/(1-\alpha)(D-2)\right)$. Whether $|Q|\rightarrow\infty$ occurs at $\tau\rightarrow\tau_*+$ or $\tau\rightarrow\tau_*-$ depends on the sign of $\zeta$: for $\zeta > 0$ it occurs at $\tau\rightarrow\tau_*-$, while for $\zeta < 0$ it would occur at $\tau\rightarrow\tau_*+$. Since $\alpha < 1$, for $\alpha \geq 0$ we have $\zeta > 0$. For $\alpha < 0$ we need to dig deeper (see below). In any event, just as for $\alpha=0$, there is a coordinate singularity at $\tau = \tau_*$, the space is geodesically incomplete, and in any physically meaningful solution we would have to continue through the $\tau = \tau_*$ point by sewing two geodesically incomplete solutions with $Q$ going to the opposite infinities as $\tau\rightarrow \tau_*$ from two different sides of $\tau_*$, just as in the $\alpha=0$ case.

{}Now that we have established that asymptotically we can only have $Q\rightarrow\mbox{const.}$, asymptotically we have three possibilities: {\em i)} ${\widetilde B}\rightarrow \mbox{const.}$, {\em ii)} ${\widetilde B}\rightarrow +\infty$, and {\em iii)} ${\widetilde B}\rightarrow -\infty$. Let us assume that asymptotically we have ${\widetilde B}\rightarrow \mbox{const.}$, so ${\widetilde B}_\tau\rightarrow 0$. Since ${\widetilde Q}_\tau\rightarrow 0$, (\ref{Q1}) implies that we must have $Q\rightarrow -1$, and then (\ref{B2}) implies that ${\widetilde B}\rightarrow 0$. We can therefore linearize (\ref{Q2}) and (\ref{B2}) in the asymptotic regime: $Q = -1 + q$, where $|q| \ll 1$ and $|{\widetilde B}|\ll 1$:
\begin{eqnarray}\label{Q3}
 &&q_\tau \approx \left[(1-2\alpha)(D-2) + 1\right]q - {2\alpha\over{1-\alpha}}\left[(1-\alpha)(D-2) + 1\right]{\widetilde B}~,\\
 &&{\widetilde B}_\tau \approx -\left[(1-2\alpha)(D-2) - 1\right]q + 2\alpha(D-2){\widetilde B}~.\label{B3}
\end{eqnarray}
So, we have a system of first order linear differential equations with constant coefficients. Then we have $q \sim q_1 \exp(\lambda\tau)$ and ${\widetilde B} \sim {\widetilde B}_1 \exp(\lambda\tau)$, where $\lambda$ can take two values given by the eigenvalues $\lambda_1$ and $\lambda_2$ of the $2\times 2$ matrix ${\cal D}$ of the coefficients in the system (\ref{Q3}) plus (\ref{B3}). From $\lambda_1 + \lambda_2 = \mbox{Tr}({\cal D}) = (D-1)$ and $\lambda_1\lambda_2 = \det({\cal D}) = 2\alpha(D-1)/(1-\alpha)$ we get
\begin{equation}
 \lambda_{1,2} = {{D-1}\over 2} \left[1 \pm \sqrt{{1 - \alpha/\alpha_c}\over{1-\alpha}}\right]~,
\end{equation}
where
\begin{equation}
 \alpha_c\equiv {{D-1}\over{D+7}}.
\end{equation}
That is, for $\alpha > \alpha_c$ such asymptotic solutions do not exist. This occurs when graviton mass is $M^2 > M_c^2$, where
\begin{equation}
 M_c^2 \equiv {M_*^2\over{1-\alpha_c}} = {1\over 4}(D-1)(D+7)H^2~.
\end{equation}
Note that these critical values $\alpha_c$ and $M_c^2$ are valid for $\beta = 1/2$ and it is natural to assume that they generally should depend on $\beta$. We derive $\alpha_c$ and $M_c^2$ for general $\beta$ in the next section.

{}For $\alpha = \alpha_c$ the eigenvalues are degenerate. For $0<\alpha <\alpha_c$ both eigenvalues are positive, so such asymptotic solutions can only occur at $\tau\rightarrow-\infty$, which is consistent with what we found above that for $\alpha > 0$, where we can only have $|Q|\rightarrow \infty$ at $\tau\rightarrow\tau_*-$. For $\alpha = 0$ one of the eigenvalues vanishes: $\lambda_2 = 0$; this is, as we found above, because in this case we have a first order differential equation for $Q$. When $\alpha < 0$, $\lambda_1$ is positive, while $\lambda_2$ is negative. So, we can have two different types of solutions, one where $Q\rightarrow -1$ at $\tau\rightarrow-\infty$ and $|Q|\rightarrow\infty$ at $\tau\rightarrow\tau_*-$, and the other where $Q\rightarrow -1$ at $\tau\rightarrow+\infty$ and $|Q|\rightarrow\infty$ at $\tau\rightarrow\tau_*+$. As we mentioned above, we must sew two geodesically incomplete solutions at $\tau = \tau_*$ to obtain a geodesically complete solution, which, as in the $\alpha=0$, case asymptotically goes to de Sitter ($Q\rightarrow-1$) on both ends.

{}We still need to check if we can have ${\widetilde B}\rightarrow +\infty$ or $-\infty$ asymptotically. In the former case from (\ref{B1}) we have ${\widetilde B}_\tau\sim \alpha(D-2)\exp(2{\widetilde B})$, whose solution behaves as $\exp(-2{\widetilde B})\sim -2\alpha\tau$, so our assumption that ${\widetilde B}\rightarrow +\infty$ does not hold. However, {\em a priori} we could have ${\widetilde B}\rightarrow -\infty$. In this case from (\ref{Q1}) and (\ref{B1}) it follows that $Q_\tau\sim [Q+1/(1-\alpha)]\left[(1-\alpha)(D-2)Q^2+1\right]$, so we invariably have $Q\rightarrow -1/(1-\alpha)$ and ${\widetilde B}_\tau\rightarrow -\alpha/(1-\alpha)$. Furthermore, for $\alpha > 0$ this can only occur at $\tau\rightarrow -\infty$. This is because in this case asymptotically we must have $q\rightarrow 0$, where $q\equiv Q+1/(1-\alpha)$, and we have $q_\tau \approx (D-1+\alpha)/(1-\alpha)q$, and the coefficient in front of $q$ is positive for $\alpha > 0$. However, since the assumption is that ${\widetilde B}\rightarrow-\infty$, this implies that for $\alpha > 0$ this could only occur as $\tau\rightarrow +\infty$ as we have ${\widetilde B}_\tau\rightarrow -\alpha/(1-\alpha)$. So, such asymptotic solutions cannot exist for $\alpha > 0$. On the other hand, for $\alpha < 0$ we can have ${\widetilde B}\rightarrow -\infty$ only at $\tau\rightarrow -\infty$, which implies that we must have $\alpha > -(D-1)$. Furthermore, we have $B = {\widetilde B} - \tau \sim -\tau/(1-\alpha)$, so $B\rightarrow +\infty$. Also, $A\sim -\tau/(1-\alpha)\rightarrow +\infty$. For the scalar curvature we have
\begin{eqnarray}
 &&R = \mu^2\left[G^{MN}{\widetilde G}_{MN} - D\right] = \nonumber\\
 &&\mu^2\left[{1\over H^2t^2}\left(e^{-2B} + (D-1)e^{-2A}\right) - D\right] \sim D\nu^2\mu^2 \exp\left({2\alpha\tau\over{1-\alpha}}\right)\rightarrow +\infty,
\end{eqnarray}
where $\nu$ is an integration constant (from $t = \exp(\tau)/\nu$). So, we have a singularity. In fact, this is a naked singularity. Thus, for the ``proper time" we have $d{\widetilde\tau} \equiv \exp(B)dt = \exp(B + \tau)d\tau/\nu = \exp({\widetilde B})d\tau/\nu$, so we have ${\widetilde \tau}\rightarrow {\widetilde \tau}_0$ as $\tau\rightarrow -\infty$, where ${\widetilde \tau}_0$ is a finite integration constant. That is, we have a true naked singularity at a finite ``proper time". So, ${\widetilde B}\rightarrow -\infty$ solutions are not physical and must be discarded.\footnote{\, Perhaps adding higher-curvature terms could smooth out this singularity.} For $\alpha <0$ this leaves us with solutions where asymptotically $Q\rightarrow -1$ and ${\widetilde B}\rightarrow 0$. Note that the only solution where asymptotically we have $Q\rightarrow -1$ for both $\tau\rightarrow -\infty$ and $\tau\rightarrow +\infty$ is the de Sitter solution itself where $Q\equiv -1$. In all other solutions asymptotically we have $Q\rightarrow -1$ at $\tau\rightarrow -\infty$ or $+\infty$ and $|Q|\rightarrow\infty$ at $\tau\rightarrow\tau_*-$ respectively $\tau\rightarrow\tau_*+$, and at $\tau = \tau_*$ we must sew two geodesically incomplete solutions into a geodesically complete solution as in the $\alpha = 0$ case.\footnote{\, These solutions are qualitatively similar to the exact solution we found in the $\alpha=0$ case and can be obtained numerically.}

\section{$\alpha_c$ for General $\beta$}

{}Let us study the full equations of motion (\ref{R.eom}) and (\ref{phi.eom}) with the metric of the form (\ref{ansatz}). Our goal in this section is to derive $\alpha_c$ for general $\beta$. To do this, let us assume that the metric $G_{MN}$ asymptotically approaches the background metric ${\widetilde G}_{MN}$, so we have $A \equiv -\ln(Ht) + {\widetilde A}$, $B \equiv -\ln(Ht) + {\widetilde B}$, and eventually we will keep only the linear terms in ${\widetilde A}$ and ${\widetilde B}$. The exact equations of motion read:
\begin{eqnarray}
 &&{{\widetilde \Lambda}\over\mu^2}\left[{\widetilde A}_{\tau\tau} + {\widetilde B}_\tau - {\widetilde A}_\tau{\widetilde B}_\tau\right] =
 -V^\prime(X)(D-1)\left[e^{2{\widetilde B}-2{\widetilde A}}-1\right],\\
 &&{{\widetilde \Lambda}\over\mu^2}\left({\widetilde A}_\tau - 1\right)^2 =
 V^\prime(X)\left[(D-1)e^{2{\widetilde B}-2{\widetilde A}}-1\right]
 +\left[V(X) - X V^\prime(X)\right]e^{2{\widetilde B}},\\
 &&(D-1)\left[e^{2{\widetilde B}-2{\widetilde A}}-1\right] + (D-1){\widetilde A}_\tau - {\widetilde B}_\tau + {V^{\prime\prime}(X)\over V^\prime(X)}~X_\tau = 0~.
\end{eqnarray}
The linearized expression for $X$ is given by $X = D - 2\left[(D-1){\widetilde A} + {\widetilde B}\right]$, so we have
\begin{eqnarray}
 && (D-1)A = (D-1)\left[\beta q + {\widetilde B}\right] - (1-\beta){\widetilde B}_\tau~,\\
 && X = D - 2\left[D{\widetilde B} + (D-1)\beta q - (1-\beta){\widetilde B}_\tau\right]~,
\end{eqnarray}
where $q\equiv{\widetilde A}_\tau$, $M_*^2/M^2\equiv 1-\alpha$, $M^2 = 2\mu V^\prime(D)$, $M_*^2 = {\widetilde\Lambda}/(D\beta-1)$, ${\widetilde \Lambda} = (D-1)(D-2)H^2$, and $\beta\equiv 1/2 - V^{\prime\prime}(D)/V^\prime(D)$. So, we have the following system:
\begin{eqnarray}
 && q_\tau = \left[(D-1) - {\alpha(D\beta-1)\over\beta(1-\beta)}\right] q - {\alpha\over{1-\alpha}}\left[{{D-1}\over{1-\beta}} -  {\alpha(D\beta-1)\over\beta(1-\beta)}\right]{\widetilde B}~,\\
 && {\widetilde B}_\tau = -\left[(D-1)-{1\over\beta}-{\alpha(D\beta-1)\over\beta(1-\beta)}\right]q + {\alpha(D\beta-1)\over\beta(1-\beta)}~{\widetilde B}~.
\end{eqnarray}
Let the matrix of the coefficients on the r.h.s. be ${\cal D}$. Then the eigenvalues of this matrix $\lambda_1$ and $\lambda_2$ satisfy the following equations:
$\lambda_1 +\lambda_2 = \mbox{Tr}({\cal D}) = (D-1)$, and $\lambda_1\lambda_2 = \det({\cal D})= \alpha(D-1)/(1-\alpha)(1-\beta)$. This implies that
\begin{equation}
 \lambda_{1,2} = {{D-1}\over 2}\left[1\pm\sqrt{{1-\alpha/\alpha_c}\over{1-\alpha}}\right]~,
\end{equation}
where
\begin{equation}\label{alpha.crit.gen}
 \alpha_c\equiv {{D-1}\over{D+{{3+\beta}\over{1-\beta}}}}~,
\end{equation}
so solutions (other than de Sitter itself) where asymptotically we have de Sitter do not exist for $\alpha > \alpha_c$, or for $M^2 > M_c^2$, where
\begin{equation}
 M_c^2 = {M_*^2\over{1-\alpha_c}} = {(D-1)(D-2)\over{4(D\beta-1)}}\left(D(1-\beta)+ 3+\beta\right)H^2~.
\end{equation}
Note that for $\beta=1$ we have $M_c^2 = (D-2)H^2 = M_*^2$~,
so there are no such asymptotic solutions for $M^2 > M_*^2$ in this case, while for $M^2\leq M_*^2$ we do have such solutions. This is the restatement of the Higuchi bound for $\beta=1$: the naive perturbative ghost instability for $M^2 < M_*^2$ translates into the fact that we have other solutions to the full non-perturbative equations of motion where the space is de Sitter only asymptotically. There is nothing ``wrong" with these solutions, in fact, perhaps they are even more interesting than de Sitter. But there is no catastrophic ``instability" such as the space-time collapsing; there is a contraction followed by an expansion with an epoch where the space appears to be (nearly) flat. Also, note that for $\beta=1/D$ we have $\alpha_c = 1/(D-1)$, albeit $M_*$ is infinite at this point, and so is $M_c$, so we have such asymptotic solutions for all values of $M$ in this extreme case.

\section{Concluding Remarks}

{}Throughout this paper we deliberately kept the space-time dimension $D$ arbitrary. This is done for two main reasons. First, more prosaically, calculations are less error-prone this way. Second, while for cosmological implications $D=4$ is the interesting case, in the event that our results may find application in string theory, it is desirable to have arbitrary $D$.

{}The non-perturbative massive solutions found in \cite{ZK.massive} in the Minkowski case are oscillatory -- the space expands and contracts eternally, in some solutions along just one dimension (such solutions were dubbed as ``cosmological strings" in \cite{ZK.massive}\footnote{\, The ``cosmological string" solutions spontaneously break spatial rotational symmetry. However, in the Minkowski case there are other, isotropic oscillating solutions as well \cite{ZK.massive}.}). Here, in the massive Sitter case, we find solutions that asymptotically in the past start as de Sitter, contract, and then expand again, with de Sitter asymptotically in the future. It would be interesting to study if there are tunneling effects such that effectively ``oscillating" solutions could be obtained semi-classically in the massive de Sitter case.\footnote{\, It would also be interesting to see if the solutions of the type we found here exist in the context of the ``ghost-free" massive gravity -- see \cite{dR} for a recent review.}

{}Finally, it would be interesting to understand if non-perturbative massive solutions we found here might have implications for or provide yet another alternative to the inflationary scenario (see, {\em e.g.}, \cite{Guth,Linde,Pad}) -- here we have no scalar fields (in the broken phase); however, ``handwavingly" one may imagine that the fake perturbative would-be ghost secretely plays the role of a scalar. It would likely require developing new non-perturbative techniques to understand this issue in detail.

\subsection*{Acknowledgments}
{}I would like to thank Olindo Corradini and Alberto Iglesias for reading an early version of the manuscript.


\end{document}